\documentclass[08pt,twocolumn,preprintnumbers,amsmath,amssymb,nofootinbib,superscriptaddress,pra]{revtex4-2}

\usepackage{amsmath,amssymb,graphicx}
\usepackage{float}
\usepackage{latexsym} 
\usepackage{amsmath,amsthm}
\usepackage{ifpdf}
\usepackage{epstopdf}
\usepackage[dvipsnames]{xcolor}
\usepackage{epsfig}
\usepackage{dcolumn}
\usepackage{bm}
\usepackage{braket}
\usepackage[symbol]{footmisc}
\usepackage{soul}
\usepackage{xcolor}
\usepackage{graphicx}

\begin{document}
\def \beq{\begin{equation}}
\def \eeq{\end{equation}}
\def \bea{\begin{eqnarray}}
\def \eea{\end{eqnarray}}
\def \bes{\begin{split}}
\def \ees{\end{split}}
\def \besu{\begin{subequations}}
\def \esu{\end{subequations}}
\def \bea{\begin{align}}
\def \eal{\end{align}}
\def \bem{\begin{displaymath}}
\def \eem{\end{displaymath}}
\def \P{\Psi}
\def \Pd{|\Psi(\boldsymbol{r})|}
\def \Pds{|\Psi^{\ast}(\boldsymbol{r})|}
\def \Po{\overline{\Psi}}
\def \bs{\boldsymbol}
\def \dert{\frac{d}{dt}}
\def \k{\ket}
\def \br{\bra}
\def \bm{\hat b^-_{\Omega}}
\def \bp{\hat b^+_{\Omega}}
\def \am{\hat a^-_{\omega}}
\def \ap{\hat a^+_{\omega}}
\def \pau {\partial_u}
\def \pav{\partial_v}
\def \paut{\partial_{\tilde u}}
\def \pavt{\partial_{\tilde v}}
\def \a{\alpha_{\Omega \omega}}
\def \b{\beta_{\Omega \omega}}

\title{Measurement of Penrose superradiance in a photon superfluid}
\author{Maria Chiara Braidotti}
\affiliation{School of Physics and Astronomy, University of Glasgow, G12 8QQ, Glasgow, United Kingdom}
\author{Radivoje Prizia}
\affiliation{School of Physics and Astronomy, University of Glasgow, G12 8QQ, Glasgow, United Kingdom}
\affiliation{Institute of Photonics and Quantum Sciences, Heriot-Watt University, EH14 4AS, Edinburgh ,United Kingdom}
\author{Calum Maitland}
\affiliation{Institute of Photonics and Quantum Sciences, Heriot-Watt University, EH14 4AS, Edinburgh ,United Kingdom}
\author{Francesco Marino}
\affiliation{CNR-Istituto Nazionale di Ottica, L.go E. Fermi 6, I-50125 Firenze, Italy}
\affiliation{INFN, Sez. di Firenze, Via Sansone 1, I-50019 Sesto Fiorentino (FI), Italy}
\author{Angus Prain}
\affiliation{School of Physics and Astronomy, University of Glasgow, G12 8QQ, Glasgow, United Kingdom}
\author{Ilya Starshynov}
\affiliation{School of Physics and Astronomy, University of Glasgow, G12 8QQ, Glasgow, United Kingdom}
\author{Niclas Westerberg}
\affiliation{School of Physics and Astronomy, University of Glasgow, G12 8QQ, Glasgow, United Kingdom}
\author{Ewan M. Wright}
\affiliation{Wyant College of Optical Sciences, University of Arizona, Tucson, Arizona 85721, USA}
\affiliation{Institute of Photonics and Quantum Sciences, Heriot-Watt University, EH14 4AS, Edinburgh ,United Kingdom}
\author{Daniele Faccio}
\affiliation{School of Physics and Astronomy, University of Glasgow, G12 8QQ, Glasgow, United Kingdom}
\affiliation{Wyant College of Optical Sciences, University of Arizona, Tucson, Arizona 85721, USA}

\email{ewan@optics.arizona.edu; daniele.faccio@glasgow.ac.uk; MariaChiara.Braidotti@glasgow.ac.uk}

\begin{abstract}
\noindent 
The superradiant amplification in the scattering from a rotating medium was first elucidated by Sir Roger Penrose over 50 years ago as a means by which particles could gain energy from rotating black holes. Despite this fundamental process being ubiquitous also in wave physics, it has only been observed once experimentally, in a water tank, and never in an astrophysical setting.
Here, we measure this amplification for a nonlinear optics experiment in the superfluid regime. In particular, by focusing a weak optical beam carrying orbital angular momentum onto the core of a strong pump vortex beam, negative norm modes are generated and trapped inside the vortex core, allowing for amplification of a reflected beam. Our experiment demonstrates amplified reflection due to a novel form of nonlinear optical four-wave mixing, whose phase-relation coincides with the Zel'dovich-Misner condition for Penrose superradiance in our photon superfluid, and unveil the role played by negative frequency modes in the process. 
\end{abstract}


\maketitle

\begin{figure}[t] 
\centering
\includegraphics[width=\columnwidth]{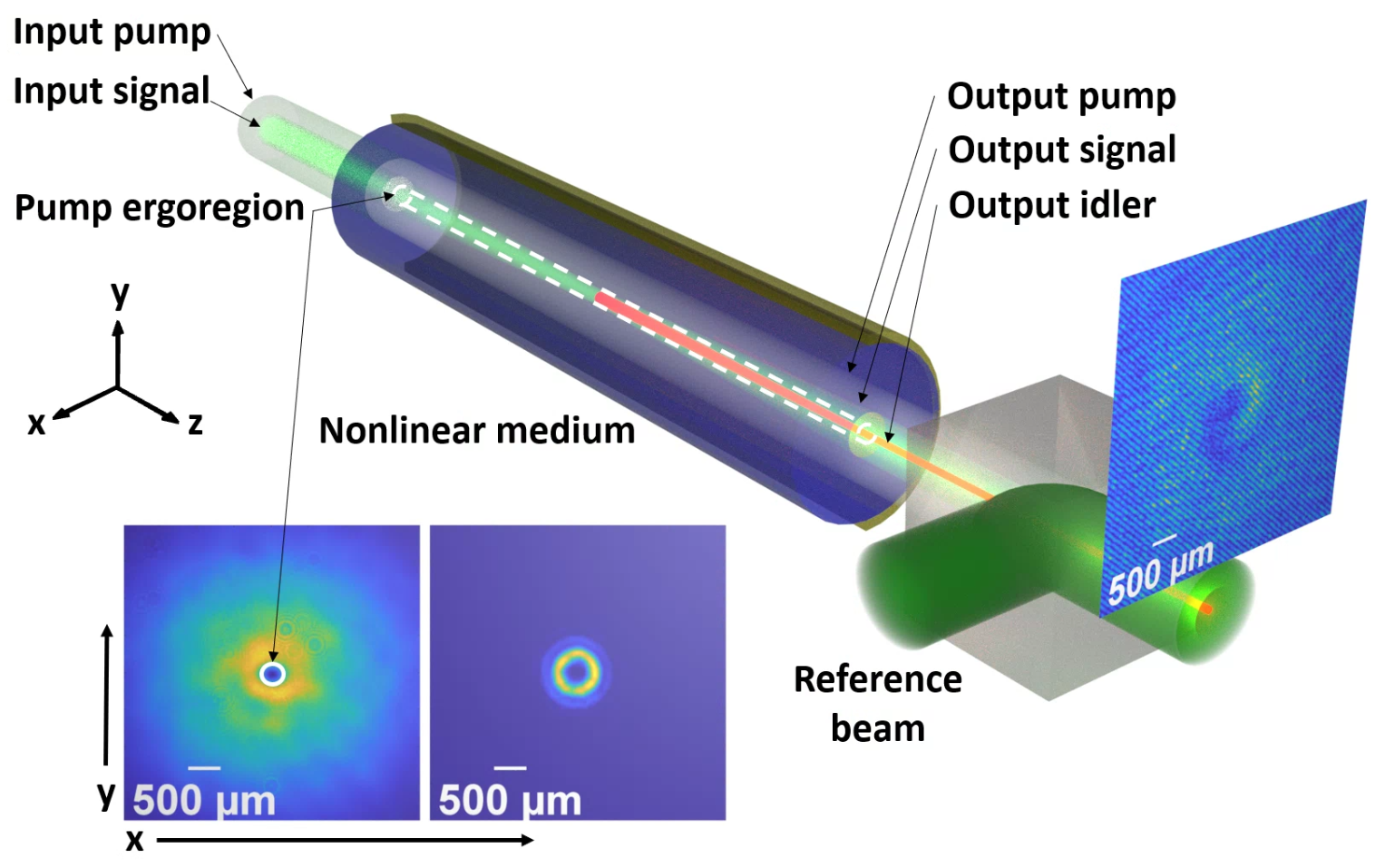}
\caption{{Illustration of the experimental   interaction geometry. The pump (grey) beam propagates inside the nonlinear sample co-propagating along the z-axis with the signal (green) beam. The dashed white line represents the ergoregion encircling the pump core. The signal beam is loosely focused onto the pump core and after the focus, if superradiance occurs, an idler (red) beam is generated and is trapped inside the ergoregion. A reference beam interferes with the total field at the output of the sample. The inset shows experimental image examples of the pump (left, $\ell=1$) and signal (right, $n=2$) beam transverse profiles at the input. The white line in the pump inset shows the ergoregion location ($r_e=118~\mu$m). The experimental parameters for the beams shown are: Pump Gaussian waist $w_{0}^{bg}\simeq1$cm, core waist $w_0^v\simeq100$~$\mu$m, and the signal core waist  $w_s^v\simeq150$~$\mu$m.}}
\label{fig:sketch}
\end{figure}

\emph{Introduction} -- The amplification via scattering from a rotating black hole -- Penrose superradiance -- derives its name after Sir Roger Penrose's 1969 prediction of the phenomenon.
He identified this process as a way to extract energy from rotating black holes, noticing that an observer at infinity \cite{ObsInf}, seeing a lump of matter splitting into two, views part of the incident co-rotating particle that becomes trapped inside the ergoregion as having negative energy, thereby allowing for the amplification of the reflected positive energy counterpart \cite{Penrose1969}.
The role of negative frequencies in rotational amplification of waves was subsequently investigated in a different setting by Zel'dovich, who showed that electromagnetic waves incident radially on a rotating metallic cylinder could be amplified if the cylinder spins fast enough to allow negative Doppler-shifted wave frequencies \cite{Zeldy1971}. Despite several proposals to observe the Zel'dovich effect for electromagnetic waves \cite{goodingReinventing2019,MC}, experimental tests have remained elusive due to the very high rotational frequencies involved \cite{Silke_sound,Ewan19,goodingDynamics2020}. To date, the only experimental observation involving a rotating solid absorber has used sound waves \cite{Marion}.\\
Astrophysical Penrose superradiance has to date not been observed with current technology, in part, due to the large distances 
between the earth and the nearest rotating black hole. However, proposals based on analogue gravity have opened up the possibility of exploring this untested gravitational phenomenon through table top experiments. 
In this context, the first experimental observation of superradiance was reported in a hydrodynamic experiment \cite{Torres2017}. In this frame, the authors measured the amplification of water waves, corresponding to the amplified positive norm modes, by a rotating vortex in a water tank. These effects are also related to the so-called over-reflection \cite{Fridman2008,Ribner1957,Miles1957,Acheson1976,Takehiro1992}.\\ 
Other analogues of astrophysical phenomena, including Hawking radiation and boson stars, have also been proposed in a variety of systems such as Bose-Einstein condensates, nonlinear optics, and hydrodynamics \cite{Liberati,FaccioBook,Garay_1,Garay_2,Schutzhold,Giovanazzi2004,Parentani2011,Steinhauer2014_1,Steinhauer2014_2,Faccio2010,Cardoso,GlorieuxPRL2021}. 
In this arena, photon superfluids have proved to be a versatile platform  \cite{Rica,Chiao1,Wan07,Conti07,Carusotto2012, Elazar2012,Carusotto2014,CarusottoLarre2015,Vocke2015,Vocke16, GlorieuxPRL2018,Vocke2018}, and recently the concept of rotational Penrose amplification has been extended to the superfluid regime, where Bogoliubov excitations act as particles in the Penrose picture \cite{Marino2008_1, Marino2009, Prain2019, Solnyshkov2019, Braidotti2020}. 
Moreover, we have recently shown that superradiance arises naturally in nonlinear optics in the superfluid regime \cite{Prain2019,Braidotti2020}. Briefly, it was shown that a wave incident at glancing angle onto the ergoregion of a rotating photon fluid undergoes a splitting into positive and negative frequency components. Here the Zel'dovich-Misner condition takes the form of a phase-relation, which if met allows for the negative frequency modes to be trapped and amplification of the positive frequency modes to occur.
 This trapping is identified by the presence of a negative Noether current inside the ergoregion \cite{Prain2019}, and allows for the calculation of the reflection and transmission coefficients for the wave incident on the ergoregion. When the superradiance condition is met, a reflectivity that is greater than one is observed, meaning that positive energy modes are amplified.\\    
In this Letter, we report measurements of Penrose rotational superradiance in nonlinear optics. Thanks to the recent findings reported in Refs.~\cite{Prain2019,Braidotti2020}, we experimentally test the nonlinear interaction of a weakly focused probe beam having Orbital Angular Momentum (OAM) and co-rotating with a strong co-propagating vortex pump beam. We demonstrate that amplification occurs only if the superradiance condition is met and idler waves, that play the role of negative energy modes, are generated and trapped inside the pump ergoregion. The presence of a negative current near the pump vortex core is a key component that underpins the physical amplification process of Penrose superradiance and had not been observed before. Photon fluids therefore allow for the study of the fundamental inner workings of Penrose superradiance, including e.g. the details and influence of negative-mode trapping and transient phenomena that are not accessible in other systems. \\
{\emph{The model}} --
We consider the nonlinear interaction between a pump beam $E_0$ and a weak signal beam $E_s$, with orbital angular momenta $\ell$ and $n$, respectively. The two fields are monochromatic with the same wavelength $\lambda$, and are co-propagating along the z-axis in a thermo-optic nonlinear medium \cite{Vocke2015,Vocke16}. The geometry of the process is depicted in Fig.~\ref{fig:sketch}. Specifically, the collimated pump vortex $E_0$ is shown as the grey beam in Fig.~\ref{fig:sketch}), and the signal beam $E_s$ (green beam in Fig.  \ref{fig:sketch}) is loosely focused onto the vortex core. The total optical field can then be written as $E=E_0+E_s+E_i$, where $E_i$ is the `idler' beam (red beam in Fig.~\ref{fig:sketch}) generated by the nonlinear interaction of the pump $E_0$ with the signal field $E_s$, and whose OAM is $q=(2\ell-n)$.  The evolution of the total field $E$ is governed in the paraxial regime by the Nonlinear Schr\"{o}dinger equation (NSE) \cite{BoydBook} 
\begin{equation} \label{fullNLS}
i\frac{\partial E}{\partial z} + \frac{1}{2k} \nabla_{\bot}^2 E + \frac{k}{n_0} \Delta n(|E|^2) E = 0,
\end{equation}  
where $n_0$ is the linear refractive index, $k=2\pi n_0/\lambda=n_0k_0$ is the wave-number, and $\nabla_{\bot}^2$ is the transverse Laplacian describing beam diffraction. Equation~(\ref{fullNLS}) is akin to the Gross-Pitaevskii equation for a two-dimensional (2D) superfluid, with $E(x,y,z)$ being the order parameter, and $z$ playing the role of time. For this photon superfluid the photon-photon repulsive interaction is mediated through the heating induced by the local intensity of the propagating beam. The specific thermo-optic medium used for the experiment is composed of a solution of methanol and a low concentration of graphene nanoflakes ($23\times 10^{-6}$g/cm$^3$), providing a weak absorption of the pump input beam to enhance the thermo-optic nonlinearity \cite{comment_on_abs}. In this system, the nonlinear refractive index perturbation $\Delta n$ in steady-state is \cite{Vocke16}
\begin{equation} \label{dn}
\Delta n(\mathbf{r}) = n_2\int R(\mathbf{r}-\mathbf{r'}) |E(\mathbf{r'})|^2 d\mathbf{r'},
\end{equation}  
where $\mathbf{r}=(x,y)$ and $n_2<0$ is the nonlinear coefficient for the defocusing nonlinearity. 
For thermo-optic nonlinearities, the non-local response function $R$ can be written as $R(\mathbf{r}) = {1\over 2\pi \sigma^2} K_0\left({|\mathbf{r}|\over \sigma}\right)$, where $K_0(s)$ is the zeroth-order modified Bessel function of the second kind, and $\sigma$ being the transverse scale of the nonlocality \cite{Faccio2016,Vocke16}.  Moreover, experiments with time gated measurements \cite{Vocke16}, performed over short times ($\sim 0.2$ seconds in our case), have verified that a strong nonlinearity with a weak nonlocality ($\sigma\sim200$ $\mu$m) can be realized before thermal diffusion has reached the steady-state (around 20 s in our case).\\
Our experiment employs short data acquisition times and we hereafter treat the photon fluid system as quasi-local \cite{Vocke16} and adopt the local theory reported in Ref.~\cite{Braidotti2020}.

\begin{figure*}[t] 
\centering
\includegraphics[width=2\columnwidth]{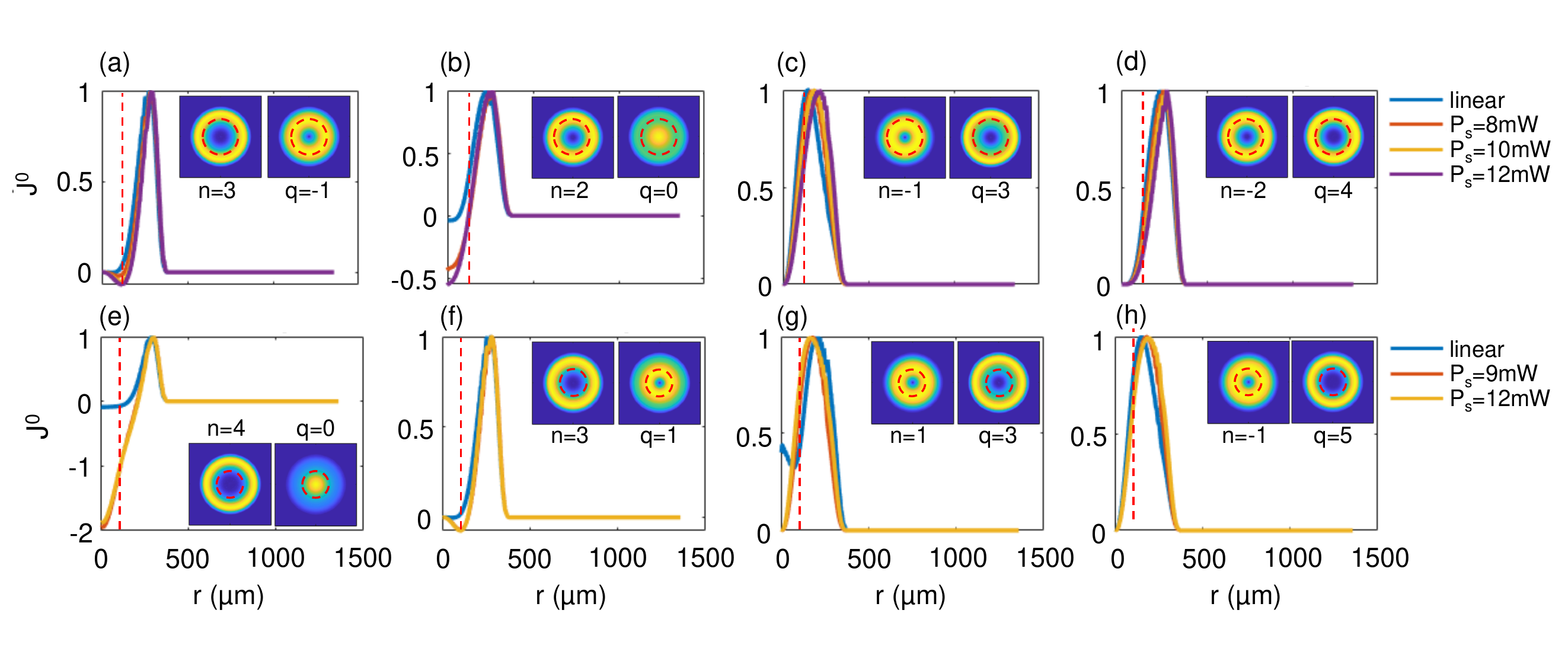}
\caption{{Current $J^0(r)$ versus radius $r$. (a-d) Pump OAM $\ell=1$ and (e-h) pump OAM $\ell=2$. Penrose superradiance conditions are satisfied in panels (a), (b), (e) and (f). (a) Signal has OAM $n=3$ and idler OAM $q=-1$. (b)  $n=2$ and $q=0$.  (c) $n=-1$ and $q=3$.  (d) $n=-2$ and $q=4$.  (e) $n=4$ and idler with OAM $q=0$. (f)  $n=3$, $q=1$. (g) $n=1$, $q=3$. (h) $n=-1$ and $q=5$. The dashed red lines in all panels indicate the location of the radius $r_e$ of the ergoregion [$r_e=118$ $\mu$m in (a-d) and $r_e=104$ $\mu$m in (e-h)]. Insets show the output intensity profiles $|E_{s,i}(r)|^2$ reconstructed from measurements verifying that idler trapping in the pump vortex core occurs for the superradiant cases (a), (b), (e) and (f). }}
\label{fig:currents}
\end{figure*}

\noindent 
All quantities are translated into the pump reference frame so that e.g. the beam longitudinal wavevectors become $\Delta K_{s,i}=k_{s,i}-\beta_\ell$, where $k_{s,i}$ are the signal and idler phonon wavevectors, and $\beta_\ell=k_0\Delta n<0$ is the pump wavevector nonlinear contribution. In a photon fluid, the phonon (i.e. transverse beam oscillation) frequency shifts, relative to the pump frequency $\omega_p=-c\beta_\ell/n_0$, are defined as $\Delta\omega_{s,i}=\omega_{s,i}-\omega_{p}=-c\Delta K_{s,i}/n_0$. A generalised phase-matching condition, $\Delta K=\Delta K_s +\Delta K_i>0$ can then be found and leads to the excitation of an idler wave that is also trapped, i.e. the idler is guided inside the core of the pump vortex beam (where the refractive index is higher than the surrounding high-intensity region)~\cite{Braidotti2020}. Importantly, this condition for efficient energy transfer to a trapped negative frequency mode was shown to be equivalent to $(\omega-m\Omega)<0$, which is exactly the Zel'dovich-Misner condition required to observe Penrose superradiance, where $\omega=(\omega_s+\omega_i)/2$, being the frequency of the Bogoliubov mode formed by the signal and idler and $\Omega=(c/n_0)|n-\ell|/(kr_e)$ is the pump transverse-plane rotation frequency \cite{Braidotti2020}. The ergoregion radius, $r_e$, is defined as the radius at which the velocity of the flow $v=|\Omega|r$ equals the speed of the photon fluid $c_s=(c/n_0)\sqrt{|\Delta n|/n_0}$ \cite{Vocke2015}.  $r_e$ can also be found as the radius of the higher order trapped idler mode, and is proportional the radius of the pump-induced waveguide potential \cite{Braidotti2020}.\\
An intuitive physical picture can be derived by noting that for plane waves, the signal and idler wave-vectors referenced to the pump beam are $\Delta K_{s,i}<0$ and their frequencies will both be positive. However, the waveguide potential will add a positive term to the idler wavevector, which can lead to $\Delta K_i>0$ and therefore $\Delta\omega_{i}<0$. Superradiance emerges then from  judicious choice of the interaction geometry i.e. from the balancing of the incoming signal that focuses onto and then diffracts away from the pump core, together with the generation of an idler beam at the focus point such that it can, under appropriate waveguiding conditions (i.e. refractive index contrast), remain trapped inside the pump vortex core.\\
In the superfluid picture, there are then three conditions to be met in order to observe Penrose superradiance. (1) First, the Zel'dovich-Misner condition $(\omega-m\Omega)<0$ should be satisfied. (2) Second, the guided idler modes should have negative frequency shifts $\Delta\omega_i<0$ to conform with the fact that they are trapped within the ergosphere. In the experiments, we explicitly verify  the conditions for $\Delta K>0$ and $\Delta K_i>0$ (i.e.  $\Delta\omega_i<0$) by numerically evaluating the signal and idler wavevectors including the contribution from the pump vortex that is modelled as a static waveguide for the idler (see SM \cite{supply}). (3) Third, since $\Omega$ is positive by definition, the Zel'dovich-Misner condition requires $m=(n-\ell)>0$ for amplification i.e. the signal OAM $n$ must be larger than the pump OAM $\ell$. This requirement is experimentally determined by the choice of input pump and signal OAM values.\\
Finally, once the experimental conditions are verified, we monitor the presence of superradiance by extracting the phonon currents. We define the reflection and transmission coefficients for scattering from the pump ergoregion using the Noether current $J^0$. The charge $N(z)$, proportional to the total energy density of the Bogoliubov modes, is a conserved quantity in this system $N(z)=\int_{0}^{\infty}J^0(r,z)rdr= \int_{0}^{\infty} \bigl( |E_S|^2 - |E_I|^2 \bigr) rdr= \mbox{const}$, where  $\partial_z J^0 = 0$ \cite{Prain2019,absorption}. The reflection and transmission coefficients are \cite{Prain2019}
\begin{align}
\label{eq:RandTcoeffs}
R_N(z) = \frac{1}{N(z)} \int_{r_e}^{\infty} \left( |E_S|^2 - |E_I|^2 \right) rdr\\
T_N(z) = \frac{1}{N(z)} \int_{0}^{r_e}      \left( |E_S|^2 - |E_I|^2 \right) rdr.
\end{align}
Since $R_N(z)+T_N(z)=1$, a {transmission $T_N<0$ implies that we must have} an amplified reflection $R_N>1$. Therefore, the generation of the trapped negative frequency (idler) mode physically underpins the superradiance process and the current $J^0(r)$ evaluated at the output can be used as a tool to prove the presence of Penrose amplification, identified as a negative current $J^0(r)<0$ inside the ergoregion (trapped idler) and $J^0(r)>0$ outside $(R_N>1)$ (reflected signal). \\
We note that the current formalism also works in the transient regime, where the signal field is still in proximity of the ergoregion and has not reached an ideal observer at infinity~\cite{Prain2019}.\\ 
\emph{Experiment} --  An outline of the experimental setup is shown in Fig.~\ref{fig:sketch} (see SM Ref.~\cite{supply} for more details). A continuous-wave Gaussian laser beam of wavelength $\lambda= 532$ nm is split into three beams: the pump beam, signal beam, and a reference beam. By using two phase masks we generate
the azimuthal phase profiles $\phi = p\theta$, with $p=\ell, n$ the OAM values for the pump and signal beams, respectively (see SM \cite{supply}).
The inset in Fig.~\ref{fig:sketch} shows the transverse intensity profiles of the pump (left) and signal (right) input beams and the corresponding beam waists are given in the caption. The two beams are then incident co-axially onto a cell of radius $1$ cm and length $13$ cm, filled with a methanol-graphene solution.\\
An additional lens is used to focus the signal beam into the pump core halfway along the medium. {The near-field intensity at the output facet of the medium and the reference plane-wave are overlapped at a small angle and imaged onto a CMOS camera. The resulting interferogram allows for the extraction of the pump+signal+idler amplitudes and phases with an off-axis digital holographic reconstruction \cite{Cuche00}.} \\
The full complex field distribution is then numerically decomposed into  Laguerre-Gauss (LG) modes so as to reconstruct the full OAM spectrum and therefore determine the precise relative compositions of the pump, signal and idler fields. 
In our experiment we considered two sets of measurements, with pump OAM $\ell=1$ and $2$ (beams core widths of 200 and 300 $\mu$m, respectively). 
%
The values of the powers (252 mW and 175 mw, respectively) are chosen in order to have a quasi-solitonic evolution of the pump vortex core. \\
Figure~\ref{fig:currents} shows the experimentally determined currents $J^0(r)$ versus radius $r$ at the output for various pump and signal OAM combinations and powers: The blue curves for low pump power ($P_p\simeq10$mW, so as to not excite any optical nonlinear effects) have $J^0(r)>0$ for all radii and superradiance is absent. For the case of pump OAM $\ell=1$, and high pump power ($P_p\simeq250$mW for OAM $\ell=1$ and $P_p\simeq175$mW for OAM $\ell=2$), shown in Figs.~\ref{fig:currents}(a-d) we analysed four signal OAMs: $n=3,2$ in Figs.~\ref{fig:currents}(a,b) that satisfy the three conditions for superradiance ($\omega-m\Omega<0$, $\Delta \omega_i<0$, $n-\ell>0$); $n=-1,-2$ in Figs.~\ref{fig:currents}(c,d) that do not satisfy the three conditions (see SM \cite{supply}). 
\begin{figure}[t!] 
\centering
\includegraphics[width=9cm]{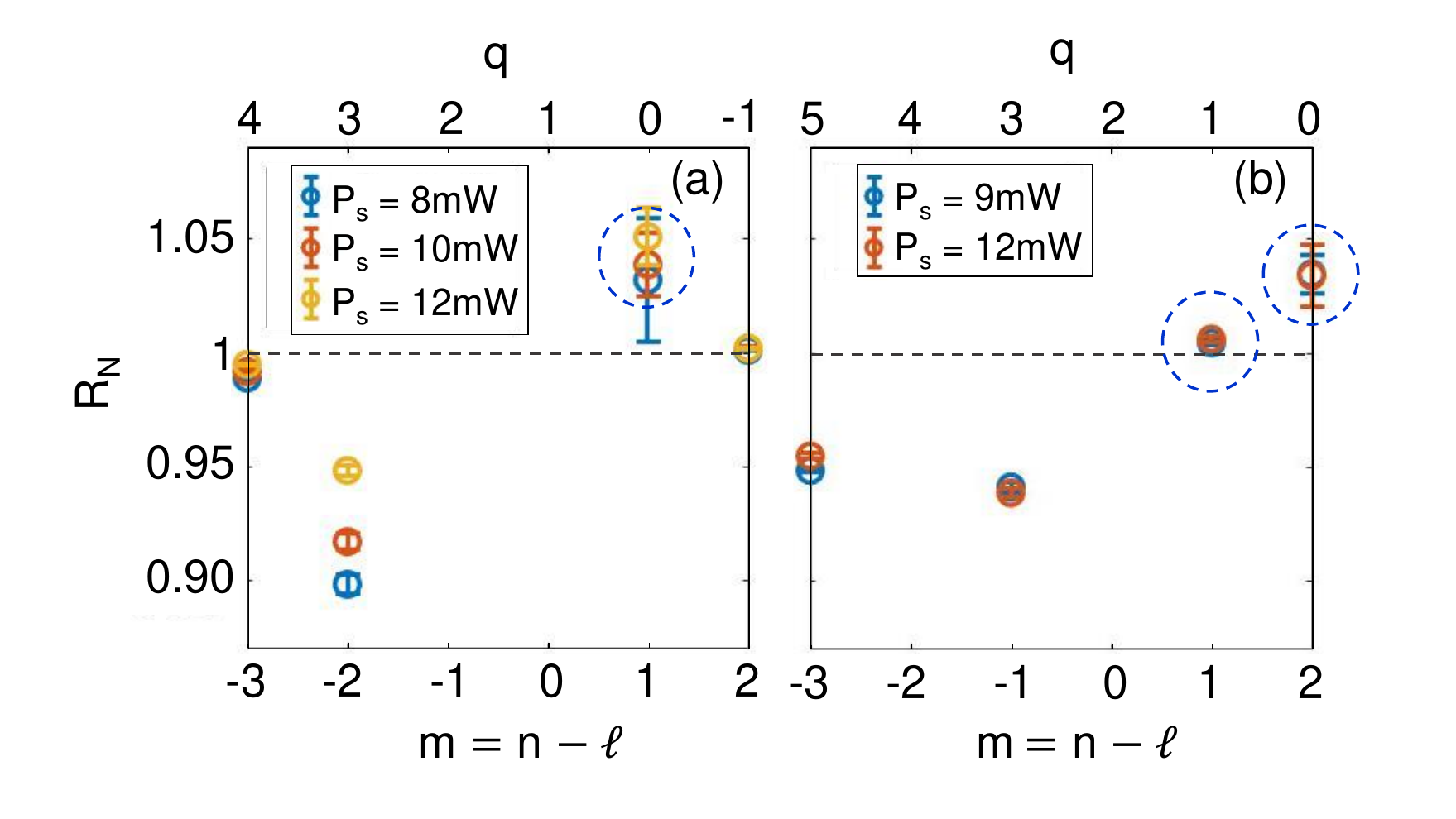}
\caption{Reflection coefficient $R_N$, calculated at the sample output $(z=13$cm), as a function of signal and pump OAM difference $m=n-\ell$ (lower axes) and idler OAM $q$ (upper axes) for: (a) pump $\ell=1$, and (b) pump $\ell=2$. Amplification with $R_N>1$ is observed for $m=n-\ell>0$. All values of $R_N$ are calculated from the average over 20 different acquisitions and the standard deviation is used to determine the error bars. Blue dashed circles indicates the configurations where superradiance conditions ($\Delta K>0$, $\Delta K_i>0$ and $m=n-\ell>0$) are satisfied (see SM \cite{supply}). }
\label{fig:refl}
\end{figure}
In particular, for Figs.~\ref{fig:currents}(a,b) we see that the current $J^0(r)$ is negative within the ergoregion (indicated in all figures by a dashed red line) and positive outside, consistent with the presence of superradiance, whereas $J^0(r)>0$ for all radii when the superradiance condition is not satisfied, Figs.~\ref{fig:currents}(c,d).\\
Figures~\ref{fig:currents}(e-h) report results for a pump OAM $\ell=2$. The two signal OAMs $n=4,3$ in Figs.~\ref{fig:currents}(e,f) display superradiance with a negative current inside the ergoregion. 
Figures~\ref{fig:currents}(g,h) are for signal OAM $n=1,-1$, which do not satisfy the Zel'dovich-Misner condition and show $J^0(r)>0$ for all $r$. \\
The insets in Fig.~\ref{fig:currents} report the corresponding  signal and idler transverse intensity profiles in the vicinity of the pump core, reconstructed from the experimental data using the LG decomposition. In particular, we observe  that the signal and idler beams can become spatially separated and that when superradiance occurs the idler beam is trapped inside the ergoregion.  This is most apparent in the examples with idler $q=0$ in Figs.~\ref{fig:currents}(b,e) where the trapped idler mode has zero OAM and is peaked in the center of the pump vortex.\\
These results therefore experimentally validate within a wide range of conditions the predicted connection between the Zel'dovich-Misner condition for Penrose superradiance and the excitation of negative norm idler modes.\\
Figure~\ref{fig:refl} shows the reflection coefficient $R_N$ calculated at the medium output for all the pump and signal OAM combinations and different signal input powers.  The reflection $R_N$ from the ergoregion is calculated using Eq.~(\ref{eq:RandTcoeffs}) as a function of $m=n-\ell$ (lower axes). The data points for which $m=n-\ell>0$ is verified together with the other two superradiance  conditions outlined above, i.e. $\omega-m\Omega<0$ and $\Delta \omega_i<0$ are circled with a dashed line (see SM \cite{supply}). As can be seen, only these points show evidence of over-reflection, $R_N>1$, and all other points have $R_N\leq1$. A maximum amplification of $5\%$ is observed when the idler OAM is $q=0$ (upper axes), consistent with the fact that the trapped idler is strongly localized in the pump vortex core.  The trend of $R_N$ with $m=n-\ell$ is also confirmed by numerical simulations (see SM \cite{supply}).  \\
%
\emph{Conclusions} --
Penrose superradiance was originally predicted as an astrophysical process in which positive energy modes are amplified from the interaction with a rotating black hole, at the expense of their negative energy counterpart, which remains trapped inside the rotating body. This concept was first tested in hydrodynamics with water waves scattering from a vortex in a bath tub. The results presented in this work show the arising of a novel process of wave mixing in nonlinear optics inspired by Penrose superradiance physics. We measure amplification of positive energy modes with OAM in the scattering with a rotating background. The trapping of the negative energy counterpart is also measured thanks to the Noether current formalism. We measure an over-reflection (reflectivity greater than one) revealing the presence of superradiance even in the transient regime. This experiment provides a novel and accessible platform for investigating Penrose superradiance, deepening our understanding of the physics at a fundamental level and for example, providing a platform for future studies investigating energy extraction from superfluid vortices. \\
\begin{acknowledgments}
\emph{Acknowledgements} --
The authors acknowledge financial support from EPSRC (UK Grant No. EP/P006078/2) and the European Union's Horizon 2020 research and innovation program, grant agreement No. 820392. NW wishes to acknowledge support from the Royal Commission for the Exhibition of 1851. The authors would like to thank Bienvenu Ndagano for the fruitful discussions and valuable suggestions throughout this project. 
\end{acknowledgments}

%


\clearpage

\onecolumngrid
{\centering{\large \bfseries Measurement of Penrose superradiance in photon superfluids:\\ Supplementary Material\par}\vspace{2ex}
	{Maria Chiara Braidotti$^{1}$, Radivoje Prizia$^{1,2}$, Calum Maitland$^2$, Francesco Marino$^{3,4}$, Angus Prain$^{1}$, Ilya Starshynov$^{1}$, Niclas Westerberg$^{1}$, Ewan M. Wright$^{2,5}$, Daniele Faccio$^{1,5}$\par}}

{\centering  \small \emph{
$^{1}$School of Physics and Astronomy, University of Glasgow, G12 8QQ, Glasgow, UK.\\ 
$^{2}$Institute of Photonics and Quantum Sciences, Heriot-Watt University, EH14 4AS, Edinburgh ,UK.\\
$^{3}$CNR-Istituto Nazionale di Ottica, L.go E. Fermi 6, I-50125 Firenze, Italy.\\
$^{4}$INFN, Sez. di Firenze, Via Sansone 1, I-50019 Sesto Fiorentino (FI), Italy.\\
$^{5}$Wyant College of Optical Sciences, University of Arizona, Tucson, Arizona 85721, USA.\\ }\par}
\smallbreak

\par\vspace{1ex}

\renewcommand{\theequation}{S\arabic{equation}}
\renewcommand{\thefigure}{S\arabic{figure}}
\setcounter{equation}{0}

\section{Experimental setup}

\begin{figure*}[h] 
\centering
\includegraphics[width=\columnwidth]{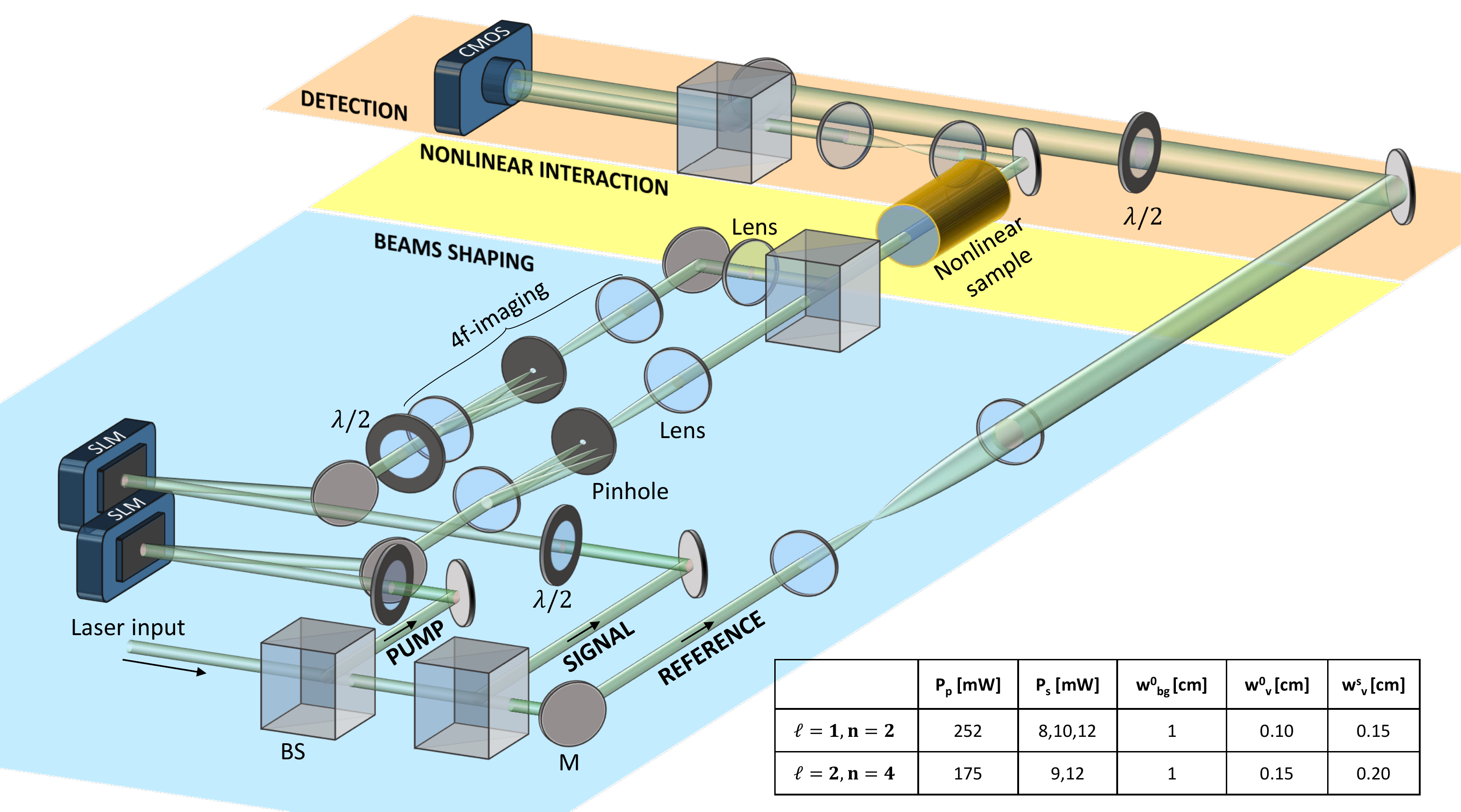}
\caption{\small{Experimental setup: a monochromatic laser beam is split into 3 components: a pump beam, a signal and a reference beam. The reference beam is expanded using a system of lenses and incident onto a CMOS camera. The pump and signal beams are shaped using two SLMs and imaged onto the front face of the nonlinear sample through a 4f-imaging system. The first diffraction order coming from the SLM is selected using a pinhole. Using a lens the signal beam is focused at the mid-plane of the medium. The pump and signal beams interact nonlinearly inside the sample. The fields at the output sample face are then imaged onto a CMOS camera, where we collect their interference with the reference beam (all beams have the same optical frequency $\lambda=532$~nm).The inset table reports the pump and signal powers, $P_p$ and $P_s$ respectively, for the two pump OAMs used $\ell = 1,2$, and two signal OAMs $n=2,4$ in the experiment, and also the values of the pump Gaussian waist $w^0_{bg}$ and pump and signal vortex core sizes, $w^0_{v}$ and $w^s_{v}$ respectively.}}
\label{fig:setup}
\end{figure*}

Figure~\ref{fig:setup} shows an illustration of the setup used: a continuous-wave laser beam with vacuum wavelength $\lambda = 532$~nm is split into three beams: a pump beam, a signal beam and a reference beam. The pump and the signal beams are shaped using two Spatial Light Modulators (SLMs), where a phase mask is designed such that the first diffracted order carries the desired phase profile: $\ell\theta$ for the pump and $m\theta $ for the signal beam. Both pump and signal beams are imaged shortly after the SLMs by a $4f-$imaging system onto the input facet of the nonlinear medium, so that the first order can be selected by a pinhole in the far-field of the first lens after each SLM.  \\
An additional lens (focal $f = 300$~mm) is placed into the signal beam path in order to weakly focus it onto the pump core into the nonlinear medium.
The medium is a cylindrical cell with a radius $W = 1$~cm and length $L = 13$~cm, filled with a solution of methanol and graphene which provides the thermal nonlinearity. The methanol absorbs a small part of the strong pump, providing a thermal defocusing nonlinearity with thermo-optic coefficient $\beta = -4 \times 10-4$~$K^{-1}$. The concentration of the graphene nanoflakes inside the methnol is $23 \times 10^{-6}$~g$/$cm$^3$ in order to increase the absorption to $25\%$ and raise the nonlinearity to $n_2 = 4.4 \times 10^{-7}$~cm$^2$/W~\cite{Vocke16}.\\
The powers used for the pump and signal beams are reported in Fig. \ref{fig:setup}, together with the beams geometrical characteristics. The values of the powers are chosen in order to have a quasi-solitonic
evolution of the pump core and ensuring linearized propagation of the signal and idler beams.
Finally, we image the near-field intensity at the output facet of the nonlinear medium onto a CMOS camera and make it interfere with a plane wave (the reference beam) (see Fig.~\ref{fig:setup}). Half-wave plates ($\lambda/2$) are placed along the beams paths in order to finely tune the beam polarizations in order to have the maximum visibility of the interference fringes. We collect with the CMOS camera the interferograms of the mixed field and the reference. By means of the off-axis digital holography technique \cite{Cuche00}, we are able to reconstruct the amplitudes and phases of the fields involved in the process. \\
The CMOS camera is gated such that it acquires the interference image $200$ ms after opening the laser shutter, with an exposure time of $20$ ms. This setting ensured that the medium has a high nonlinearity for superfluidity with a small nonlocal length, $\sigma \sim 166~\mu$m~\cite{Vocke16}.

\section{Data Analysis}
We collect the interference images of the plane wave reference beam with the field with the aim of reconstructing the field in phase and amplitude. The interference images are analysed using the Off-Axis Digital Holography technique \cite{Cuche00}: by performing a Fourier Transform on this interference profile, in $k$-space we observe the positive and negative frequency components of the field $E$ at $(k_x, k_y)$ and $(-k_x, -k_y)$, respectively. By selecting one of these two lobes and inverse Fourier transforming it, we obtain the complex field $E(x,y) = \mathcal{E}(x,y) e^{i\phi(x,y)}$, where $\mathcal{E}(x,y)$ and $\phi(x,y)$ are the retrieved amplitude and phase profiles of the target beam, respectively.\\
Through a Laguerre-Gauss (LG) decomposition of $E(x,y)$ it is then possible to distinguish the OAM components in the field, identifying the pump, signal and idler beams. We define the LG weights as 
$$w_{j,p} = \int \int dxdy L_{j,p}(x,y) E^*(x,y,z),$$ 
where $L_{j,p}$ are the Laguerre-Gauss modes. Each weight is a complex number giving the amplitude of a given LG mode in the $E$ field. 
From $w_{j,p}$ it is possible to reconstruct the $j-$components of the experimental field by the transformation $$E^j(x,y) = \int dp w_{j,p} L_{j,p}(x,y),$$
thereby finding the form of the field for each OAM $j$ component. 
In this way we have access to the $(x,y)$ dependent phase and amplitude profiles of the pump $E^{(\ell)}_0$, signal $E^{(n)}_s$ and idler $E^{(q)}_i$ fields.\\
As stated in the manuscript we chose different OAMs combinations for pump and signal in order to fully characterize the superradiance process. In particular we chose 2 pump OAMs, being $\ell=1,2$. For these two pumps, we chose 2 signal OAMs $n$ that satisfy the Penrose condition and 2 that do not. \\
For each combination of pump and signal OAMs we collect $N=20$ interferograms from which we reconstruct 20 field sets in order to have good statistics. 
From the $N-$sets of signal and idler fields it is possible to calculate the conserved charge from the current $J^0$ as defined in the manuscript, i.e.
$$
N(z)= \int \int J^0dxdy = \int \int (|E^{(n)}_s|^2-|E^{(q)}_i|^2)dxdy
$$
The current plotted in Fig 2 of the manuscript is the averaged current over the 20 values obtained. \\
Furthermore, the value of the reflection coefficient $R_N$ in Fig. 3 of the main manuscript is obtained by averaging over the $20$ values of $R_N$ obtained from Eq. (3) in the manuscript. The error bar is given by the standard deviation over the 20 measurements. 

\section{Evaluation of superradiance conditions}
As outlined in the main text we have three conditions for superradiant amplification to occur, namely:
\begin{itemize}
    \item $\Delta K>0$, i.e. $\Delta\omega<0$, which can be mapped to the Zel'dovich-Misner condition $\omega-m\Omega<0$
    \item $\Delta K_i>0$ or equivalently $\Delta\omega_i<0$, i.e. the idler mode should have negative frequency and be trapped inside the ergoregion, as in the Penrose picture
    \item $m=n-\ell>0$ (as a result of $\Omega>0$ in the first condition) i.e. the signal OAM $n$ has to be larger than the pump OAM $\ell$. 
\end{itemize}
In order to verify the first two conditions, we numerically evaluate the phase-matching condition $\Delta K\propto-\Delta\omega>0$ and the idler wave vector $\Delta K_{i}\propto-\Delta\omega_{i}>0$, by calculating the idler modes that can be guided inside the pump core guiding potential. In the present calculation the wave-guiding potential is approximated as local, and to be invariant along the propagation axis $z$. This approximation, despite being quite strong, is justified experimentally as we purposely chose the pump power to have a quasi-solitonic evolution and we performed time resolved measurements which guarantee a quasi-local wave-guiding potential, the nonlocality being small $\sigma\simeq200~\mu$m. In photon fluids the nonlocal effects are negligible if the nonlocality $\sigma$ is smaller than the so-called healing length $\xi=\lambda/\sqrt(4n_0|n_2|I_0)$~ \cite{Vocke16}. In our system, the healing length is approximately $600~\mu$m, being $n_0=1.32$, $\lambda=532$nm and $I_0=P_p/w_0^2$ (see table in Fig. \ref{fig:setup}).\\
We write the signal and idler equations of propagation as \cite{Braidotti2020},
\begin{eqnarray} 
\frac{\partial E_s}{\partial z} &=& i\frac{1}{2k} \nabla_{\bot}^2 E_s + i2\beta_\ell u_\ell^2(r)E_s - i\beta_\ell E_s + i\beta_\ell u_\ell^2(r)E_i^*,\label{sigNLS}\\
\frac{\partial E_i}{\partial z} &=&  i\frac{1}{2k} \nabla_{\bot}^2 E_i + i2\beta_\ell \left(u_\ell^2(r)-1\right)E_i + i\beta_\ell E_i + i\beta_\ell u_\ell^2(r)E_s^*,\label{ildNLS}
\end{eqnarray}  
where $u_\ell(r)$ is the pump vortex profile with core size $r_\ell$. In Eq. (\ref{ildNLS})  the second term on the right-hand-side is the waveguiding term. By neglecting the source term in Eqs. (\ref{sigNLS}-\ref{ildNLS}), last term on the right-hand-side, we can write the signal and idler as \cite{Braidotti2020} 
\begin{eqnarray} 
& E_s(r,z)\propto c_s(z)V(r)e^{i(2\beta_\ell\Gamma_n(0)-\beta_\ell)z},\\
& E_i(r,z)\propto c_i(z)U(r)e^{i(\beta_\ell+\Lambda_q)z},\label{idl}
\end{eqnarray} 
with signal and idler wave vectors given by \cite{Braidotti2020} 
\begin{eqnarray} 
& \Delta K_s = 2\beta_\ell\Gamma_n(0)-\beta_\ell,\\
& \Delta K_i=\beta_\ell+\Lambda_q,
\end{eqnarray} 
Here $\Gamma_n(z)=\int_0^{\infty} 2\pi r dr |V(r,z)|^2u_\ell^2(r)$ is the phase variation due to the overlap between the signal and the pump while the signal is focusing onto the vortex core along the propagation, and $\Lambda_q$ is the idler eigenvalue (see below) such that the idler gets trapped if $\Lambda_q>0$.\\
\begin{table}[h!]
\centering
\begin{tabular}{| l | c | c | c | c | c | c | c |} 
 \hline
 OAMs                        & $m=n-\ell>0$ & $\omega-m\Omega<0$ & $\Delta\omega_i<0$ & $n_g$ & $R_{BPM}$ & $R_{NLS}$ & $R_{N}$ (at $P_s=8,10,12$ mW) \\ 
 \hline
 \hline 
 $\ell=1$, $n=3$, $q=-1$        &    yes     &    no     &     no   &   2   & 1 & 1  & 1, 1, 1 \\ 
 \hline
 ${\mathbf{\ell=1,  {n}=2, q=0 }}$   &  \bf{yes}  & \bf{yes} & \bf{yes} & 2 & \bf{1.19}  & \bf{1.85}  & \bf{1.03, 1.04, 1.05} \\ 
 \hline
 $\ell=1$, $n=-1$, $q=3$        &     no     &    no     &    no    &   0   & 1  & 0.92  & 0.90, 0.92, 0.95 \\ 
 \hline  
 $\ell=1$, $n=-2$, $q=4$        &     no     &   no      &    no    &   0   & 1  & 1  & 1, 1, 1 \\ 
  \hline
  \hline
    \hline
  OAMs                        & $m=n-\ell>0$ & $\omega-m\Omega<0$ & $\Delta\omega_i<0$ & $n_g$ & $R_{BPM}$ & $R_{NLS}$ & $R_{N}$ (at $P_s=9,12$ mW) \\ 
  \hline
  \hline
 ${\mathbf{\ell=2, n=4, q=0}}$  &  \bf{yes}  &  \bf{yes} &  \bf{yes}   &   4   & \bf{1.7}  & \bf{1.1}  & \bf{1.03, 1.03} \\
 \hline
 ${\mathbf{\ell=2, n=3, q=1}}$  &  \bf{yes}  &  \bf{yes} &  \bf{yes}   &   3   & \bf{1.4}  &\bf{1}  & \bf{1, 1.01} \\ 
 \hline
 $\ell=2$, $n=1$, $q=3$         &   no       &     yes   &   no        &   2   & 0.7  & 0.4  & 0.94, 0,94 \\  
 \hline
 $\ell=2$, $n=-1$, $q=5$        &   no       &    no     &     no      &   1   & 1  & 0.5  & 0.95, 0.95 \\ 
 \hline
\end{tabular}
\caption{Summary of superradiance conditions in the various OAM configurations used in the experiment. When the three conditions ($\omega-m\Omega<0$, $\Delta \omega_i<0$ and $m=n-\ell>0$) or equivalently ($\Delta K>0$, $\Delta K_i>0$ and $m=n-\ell>0$) are satisfied the reflection coefficient $R$ results greater than 1 $(R>1)$, while it is not if any of the three conditions does not hold. $n_g$ is the number of idler guided modes, i.e. the number of modes with $\Lambda_q>0$. The cases in which all three SR conditions are verified are indicated in bold font. The three reflection coefficients are obtained by the local and "solitonic"-pump BPM for the signal and idler $(R_{BPM})$, by the full NLS numerical simulation $(R_{NLS})$ (see Sec. IV for more details) and by experimental measurements $(R_{N})$.}
\label{table1}
\end{table}

\noindent In order to verify the presence of trapped idler modes in the presence of the cross-phase modulation induced waveguide supplied by the pump vortex core, we compute the spectrum of the idler modes by solving the eigenequation
\begin{equation}
{1\over 4} \left [ {d^2\over dr^2} + {1\over r}{d\over dr} -{q^2\over r^2} \right ] U(r) + \Delta n(r)U(r) =  \Lambda_q U(r)
\label{eq:idl}
\end{equation}
with $\Delta n(r)=2\beta_\ell (u_\ell^2(r)-1)$ the waveguiding potential, with eigenvalue $\Lambda_q$ for a idler with OAM $q$, and $U_q(r)$ the corresponding eigenmode.  Equation (\ref{eq:idl}) is solved as a matrix eigenvalue problem by discretizing the radius spatial coordinate $r$ and writing the spatial derivatives as
\begin{equation}
{d^2 U\over dr^2}  \rightarrow {U_{j+1} -2U_j + U_{j-1}\over dr^2} , \quad {d U\over dr}  \rightarrow {U_{j+1} - U_{j-1}\over 2dr}. 
\end{equation}
Then the eigenequation for the idler modes may be turned into a matrix equation $M\bar U=\Lambda_q \bar U$, with $M$ a tridiagonal matrix. If $\Lambda_q>0$, the mode can get trapped inside the waveguiding potential and we denote $n_g$ as the number of eignemodes with positive eigenvalue $\Lambda_q$.\\
Table \ref{table1} reports the results of the phononic frequency signs from the idler mode computations and shows that when all the three conditions above are satisfied the reflectivity is larger than unity (see last columns). 
More specifically, the reflectivity $R_{BPM}$ reported in Table I is calculated using the beam propagation method (BPM) to solve Eqs.~(\ref{sigNLS}-\ref{ildNLS})) with the same parameters as in the idler mode computation, thus with a local nonlinearity and static pump evolution. The trend in reflectivity $R_{BPM}$ is found to be the same as that found experimentally $R_{N}$ (see last column of Table \ref{table1} and Fig. 3 of the paper) even if the absolute values are different. In order to further validate our analysis we performed a full numerical simulation with nonlocal nonlinearity and propagating pump beam, described in the section below (Sec. IV) and calculated the corresponding reflectivity $R_{NLS}$, reported in table \ref{table1}. We observe that the trend is confirmed also in this case, verifying the experimental results. The higher values of gain and absorption found from the numerical simulations with respect to experiments are due to the absence of experimental noise in the simulations. In particular, the alignment between pump and signal is experimentally very critical both for having phase-matching and for a correct estimation of the components in the field $E$. A small misalignment, i.e. the presence of a tilt, both ruins the phase-matching and furthermore, spreads a single OAM contribution onto the neighbours OAMs hindering the exact reconstruction of the field components. The main contribution to the lower experimental amplification values compared to the numerical values is due to a strong sensitivity to input alignment of the pump and signal beams. Indeed, in the experiments we noted that observing $R>1$ required a very precise alignment of the pump and signal beams, such that both beams are well centred and also co-propagating along the same exact z-axis (i.e. such that there is no tilt in their relative propagation directions). However, the two beams also follow different path trajectories in the setup before entering the nonlinear media - this can lead to small fluctuations in the beam alignment due to different and unavoidable fluctuations on the two paths (e.g. weak air turbulence or vibrations), which then in turn leads to large fluctuations in the measurements as $R$ quickly drops to 1 with small misalignment. 

\section{Numerical simulations}
We test the experimental conditions by numerically simulating the Nonlinear Schr\"{o}dinger equation (NSE) \cite{BoydBook} employed in the manuscript
\begin{equation} \label{fullNLS}
i\frac{\partial E}{\partial z} + \frac{1}{2k} \nabla_{\bot}^2 E + \frac{k}{n_0} \Delta n[|E|^2] E = -\frac{i\alpha}{2}E,
\end{equation}  
with 
\begin{equation} \label{dn}
\Delta n = n_2\int R(\mathbf{r}-\mathbf{r'}) |E(\mathbf{r'})|^2 d\mathbf{r'}.
\end{equation}  
We included the absorption, given by the coefficient $\alpha$, which is induced on the pump beam by the presence of the graphene flakes in the nonlinear medium ($\alpha=2$~m$^{-1}$), in order to  verify the persistence of the physics of the Penrose superradiance.
The background pump beam is generated by an incident vortex beam 
$$
E_0(r,z=0) = U_0 e^{-r^2/w_{bg}^2} \tanh^{|\ell|}(r/w_v) e^{i\ell\theta}
$$
where $\ell$ is the vortex OAM and $U_0$ controls the pump power. This form of the pump as well as the parameter used in the numerics, reported in the table \ref{table2}, are chosen in order to have a simulation as close as possible to the experimental configuration (see also Fig. 1 in the manuscript).\\
%
%
The signal field $E_s$ at the center of the medium is chosen as a Laguerre-Gauss beam with OAM $n$
$$
E_s(r,z=L/2) = U_s \left(\frac{r}{w_v}\right)^{|n|} e^{-r^2/w_v^2} e^{in\theta},
$$
where $U_s$ controls the signal power.  This focused version of the signal field can be propagated backwards linearly to the input plane at $z=0$ to determine the initially focused signal beam.
The idler beam is chosen to have zero amplitude at the input.\\
\begin{table}[h!]
\centering
\begin{tabular}{| c | c | c | c | c |} 
 \hline
 pump OAM & $P_p$ [mW] & $P_s$ [mW] &  $w_{bg}$ [cm] & $w_{v}$ [cm] \\ 
 \hline
 $\ell =1$ & 252 & 5 & 1& 0.08  \\ 
 \hline
 $\ell =2$ & 175 & 1 & 1 & 0.06 \\
 \hline
\end{tabular}
\caption{Pump and signal powers, $P_p$ and $P_s$ respectively, for the two pump OAMs used $\ell = 1,2$ in the simulations. Values of the Gaussian background waist $w_{bg}$ and of the vortex core waist $w_v$ of both pump and signal initial beams.}
\label{table2}
\end{table}
%
We compute the reflection coefficient after propagating the incident beam, signal plus pump, over the distance $z=13$~cm as in the experiment. We find the same trend as that found in the experiment reported in Fig. 3 of the manuscript which we report here as well (see Table \ref{table1}).
 The maximum amplification in the numerical simulations results are also reached when the idler OAM is $q=0$ and is of $\sim85\%$. When superradiance conditions are not met there is a maximum absorption (of the signal beam inside the pump ergoregion) of $60\%$. 
 As said before, the numerical values of gain and absorption are higher than the experimental ones due to the presence of noise in the experiment which contributes to both ruining the phase-matching and furthermore, preventing the exact analysis of the fields OAM contributions.\\ 
Nonetheless, numerical simulations confirm the results in the transient regime, showing an amplification $R_{NLS}>1$ even if the signal field has not reached the asymptotic observer. 


\begin{thebibliography}{50}
\bibitem [{Obs()}]{ObsInf}%
  \BibitemOpen
  \href@noop {} {}\bibinfo {note} {As observer at infinity we mean in a
  reference frame compatible with that of a Minkowskian observer far from the
  black hole.}\BibitemShut {Stop}%
\end{thebibliography}
%

\end{document}